\documentclass[%
reprint,
superscriptaddress,
 amsmath,amssymb,
 aps,
prr,
]{revtex4-1}

\usepackage{comment} 
\usepackage{graphicx}
\usepackage{epstopdf}
\usepackage{textcomp}
\usepackage{dcolumn}
\usepackage{bm}
\usepackage{gensymb}
\usepackage{upgreek}
\usepackage[usenames, dvipsnames]{color}
\usepackage{mathtools}
\usepackage{amsmath}
\usepackage{physics}
\usepackage{esvect}
\usepackage[mathscr]{euscript}

\newcommand{\BaNi}{BaNi$_2$As$_2$}
\newcommand{\SrNi}{SrNi$_2$As$_2$}
\newcommand{\BaFe}{BaFe$_2$As$_2$}
\newcommand{\BaSr}{Ba$_{1-x}$Sr$_{x}$Ni$_2$As$_2$}

\newcommand{\BaSrLate}{Ba$_{0.35}$Sr$_{0.65}$Ni$_2$As$_2$}
\newcommand{\Tc}{$T_c$}

\begin{document}

\preprint{APS/123-QED}

\title{Absence of precursor incommensurate charge order in 
electronic nematic 
\BaSrLate}

\author{John Collini}
\affiliation{Maryland Quantum Materials Center, Department of Physics, University of Maryland, College Park, Maryland 20742, USA}

\author{Sangjun Lee}
\affiliation{Department of Physics and Materials Research Laboratory, University of Illinois, Urbana, Illinois 61801, USA}

\author{Stella X.-L. Sun}
\affiliation{Department of Physics and Materials Research Laboratory, University of Illinois, Urbana, Illinois 61801, USA}

\author{Chris Eckberg}
\affiliation{Maryland Quantum Materials Center, Department of Physics, University of Maryland, College Park, Maryland 20742, USA}

\author{Daniel J. Campbell}
\affiliation{Maryland Quantum Materials Center, Department of Physics, University of Maryland, College Park, Maryland 20742, USA}


\author{Peter Abbamonte}
\affiliation{Department of Physics and Materials Research Laboratory, University of Illinois, Urbana, Illinois 61801, USA}

\author{Johnpierre Paglione}
\email{paglione@umd.edu}
\affiliation{Maryland Quantum Materials Center, Department of Physics, University of Maryland, College Park, Maryland 20742, USA}
\affiliation{Canadian Institute for Advanced Research, Toronto, Ontario M5G 1Z8, Canada}

\date{\today}

\begin{abstract}
Recent discoveries of charge order and electronic nematic order in the iron-based superconductors and cuprates have pointed towards the possibility of nematic and charge fluctuations playing a role in the enhancement of superconductivity. The \BaSr~system, closely related in structure to the \BaFe~system, has recently been shown to exhibit both types of ordering without the presence of any magnetic order. We report single crystal X-ray diffraction experiments on \BaSrLate, providing evidence that the previously reported incommensurate charge order with wavevector $(0,0.28,0)_{tet}$ in the tetragonal state of \BaNi~vanishes by 65\% Sr substitution together with nematic order. Our measurements suggest that the nematic and incommensurate charge orders are closely tied in the tetragonal state, and show that the $(0,0.33,0)_{tri}$ charge ordering in the triclinic phase of \BaNi~evolves to become  $(0,0.5,0)_{tri}$ charge ordering at $x$=0.65 before vanishing at $x$=0.71.   

\end{abstract}

\maketitle


High temperature superconductivity in the cuprate \cite{Cuprates_Science_Orenstein,Cuprates_Nature_Kiemer} and iron-pnictide families \cite{Fe-Sc_Nature_Paglione,Fe-Sc_AdvPhys_Johnston,Fe-Sc_RevModPhys_Stewart} has prompted wide research efforts aiming to uncover the origins of their unconventional pairing mechanisms. Although long range magnetic fluctuations have long been suggested as responsible for the pairing, recent research into electronically driven nematicity in the iron-pnictides \cite{Fisher_FeNumatic,Fernandes_Numatic_Review} as well as charge order and nematic order in the cuprates \cite{CDW_Cuprate_Discovery_first,CDW_Cuprate_Discovery_second,CDW_Cuprate_Discovery_third,CDW_Cuprate_Discovery_fourth,Cuprates_Nematic_Kivelson,Cuprates_Nematic_Auvray,Cuprates_Nematic_Liu} suggests that these electronic degrees of freedom may also play important roles in stabilizing the superconducting phases in these systems. 

Theoretical work has shown that fluctuations associated with an electronically nematic quantum critical phase can enhance superconducting phases with few requirements \cite{NematicTheory_Lederner,NematicTheory_Lederner2}. The proximity of magnetism in the cuprates and iron-pnictides prevents a straightforward study investigating the potential enhancement effects of electronic nematic fluctuations to superconductivity. Our recent studies of the tunable superconducting pnictide material \BaSr~revealed a strong sixfold enhancement of \Tc~from 0.6 K to 3.5 K in proximity to an increase of nematic fluctuations, suggesting strong evidence of a pairing enhancement \cite{Eckberg_nematic}. Furthermore, with evidence of charge \cite{Lee_BANI_CDW} and nematic \cite{Eckberg_nematic} orders in proximity to the enhancement, \BaSr~is positioned as a a good candidate for exploring the interplay of charge and nematic degrees of freedom in the absence of magnetism within superconducting systems. 

The parent compound \BaNi~is tetragonal and isostructual to its famous iron-based counterpart, \BaFe~at room temperature. Unlike the latter, \BaNi~undergoes a first-order tetragonal to triclinic structural phase transition at $T_s=135$ K. Additionally, neutron measurements of \BaNi~have shown no evidence of a magnetic structure in its low temperature phase or anywhere else \cite{Kothapalli_BANI_Neutron}. The tetragonal and triclinic phases of \BaSr~are denoted by sets of distinct Bragg peaks that index to the space groups of $I4/mmm$ and $P1$ respectively. Here we use $(H,K,L)_{tet}$ and $(H,K,L)_{tri}$ separately to describe positions in momentum space for each phase. X-ray measurements have revealed a bi-driectional incommensurate charge density wave (IC-CDW) onsetting just above $T_s$ at $T_{IC}=148$ K at a wavevector of $Q_{tet}=0.28$ in a ``$4Q$'' state in the $ab$ plane \cite{Lee_Paper2}. At $T_s$, the incommensurate CDW vanishes and gives way to a unidirectional commensurate CDW (C-CDW1) at wavevector $(0,0.33,0)_{tri}$ in the triclinic phase \cite{Lee_BANI_CDW}. 
\BaNi~also becomes superconducting at $T_c=0.7$ K \cite{Ronning_BaNiAs-SCPaper}, and thermal conductivity measurements suggest that this superconducting state is fully gapped \cite{Kurita_BaNiThermal_SC}. The other end member, \SrNi, shows no evidence for a structural distortion or magnetic order, but also superconducts below $T_c=0.62$ K \cite{Bauer_SRNI_SC}. Isovalent substitution of Sr for Ba in \BaSr~ has been shown to suppress $T_s$ toward absolute zero temperature and enhance $T_c$ up to a maximum value of 3.5 K at $x$=0.71 \cite{Eckberg_nematic}. 

Elastoresistivity measurements of \BaSr\ probing the $B_{1g}$ channel, corresponding to the symmetry-breaking strain along the [100] and [010] tetragonal crystallographic directions, have revealed a large nematic susceptibility throughout the range of Sr substitution \cite{Eckberg_nematic}. In addition, this experiment also revealed striking non-reversible hysteretic behavior in the nematic response just above the triclinic distortion between $x$=0 and $x$=0.5, implying the presence of an ordered electronic nematic phase that coexists with IC charge order in the tetragonal structural phase. In \BaNi~at temperatures just above $T_s$, the area of hysteresis in elastoresistance from nematic order sharply increases with a profile that matches the sharp intensity growth of the IC-CDW. With increasing Sr content, this nematic order begins to fade  until a crossover to electronically driven nematic fluctuations occurs in the intermediate region around $x$=0.5. The correlation between the $Q_{tet}=0.28$ charge order and nematic order, in both scaled intensity and crystallographic direction, suggests that the two phases are closely linked. In this work, we provide evidence for the vanishing of the incommensurate charge order that lives above the triclinic distortion at $x$=0.65, matching the disappearance of nematic order at the same concentration. The absence of this incommensurate charge order in a region void of nematic order further supports the notion that these two phases are linked.



Single crystal X-ray diffraction measurements were carried out using a Xenocs GeniX 3D Mo $K_{\alpha}$ (17.4 keV) source which delivers $2.5\cross 10^7$ photons per second with a beam spot of 130$ \mu$m. The sample was cooled using a closed-cycle cyrostat to reach a base temperature of 11 K. The sample was kept inside a Be dome, used for vacuum and radiation shielding. Sample motion was performed using a Huber four-circle diffractometer and X-ray detection was captured using a Mar345 image plate to allow for 3D mapping of momentum space of with a resolution of $\Delta q=0.01$ to 0.08 \AA$^{-1}$ depending on the cut\cite{Lee_BANI_CDW}. Single crystals of \BaSr~were grown using a NiAs self flux solution technique\cite{Ronning_BaNiAs-SCPaper}. Resistivity measurements were performing using a Quantum Design Physical Property Measurement System (PPMS). 

\begin{figure}
    \centering
    \includegraphics[width=0.47\textwidth]{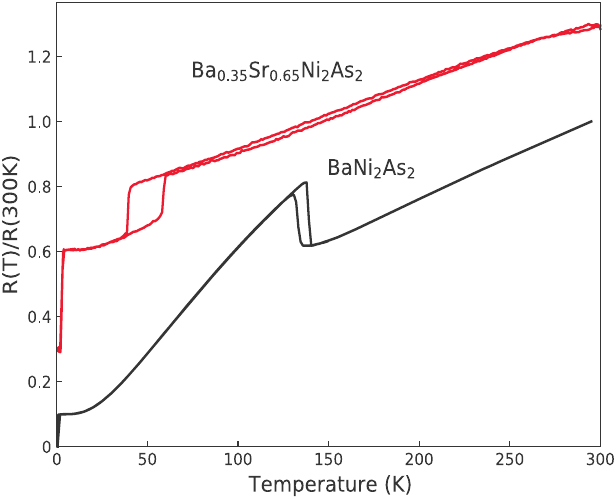}
    \caption{Resistance of \BaNi~and \BaSrLate~normalized to 300 K data. \BaSrLate~has been offset from \BaNi. The triclinic distortion in \BaNi~can be seen in resistance at 135 K on cooling and at 138K on warming and in \BaSrLate~at 40 K on cooling and at 58K on warming. The \Tc~of \BaNi~and \BaSrLate~is measured at 0.6 K and 3 K respectfully.}
    \label{fig:Transport}
\end{figure}

Charge order in the \BaSr~ system develops in a complex manner, with different phases coexisting \cite{Lee_Paper2}. Just as nematic fluctuations undergo changes in character across the phase diagram, so too does the charge order.
\BaNi~develops a complex ``$4Q$" bi-directional IC-CDW with a wavevector at $(0.28,0.28,0)_{tet}$ that onsets at 148 K \cite{Lee_Paper2}, distinctly above the 
first order triclinic transition (c.f. Fig.~\ref{fig:Transport}). The IC-CDW order grows in X-ray intensity as temperature is lowered until the structural transition occurs, where this order abruptly disappears and a new unidirectional commensurate charge (C-CDW1) order forms at $(0,0.33,0)_{tri}$ in the triclinic phase and is maintained down to base temperature \cite{Lee_BANI_CDW}. For increasing Sr concentrations on the order of $x$=0.4 to $x$=0.5, the IC-CDW's window of existence above the triclinic order narrows closer to the triclinic onset temperature \cite{Lee_Paper2}. Additionally in the triclinic phase, a new unidirectional commensurate charge order forms at $(0,0.5,0)_{tri}$ (C-CDW2) about 20 K below the triclinic onset \cite{Lee_Paper2}. For a narrow range of Sr concentration, about $x$=0.4, both C-CDW1 and C-CDW2 coexist down to base temperature with C-CDW2 being largely dominant in scattering intensity \cite{Lee_Paper2}.

\begin{figure}
    \centering
    \includegraphics[width=0.47\textwidth]{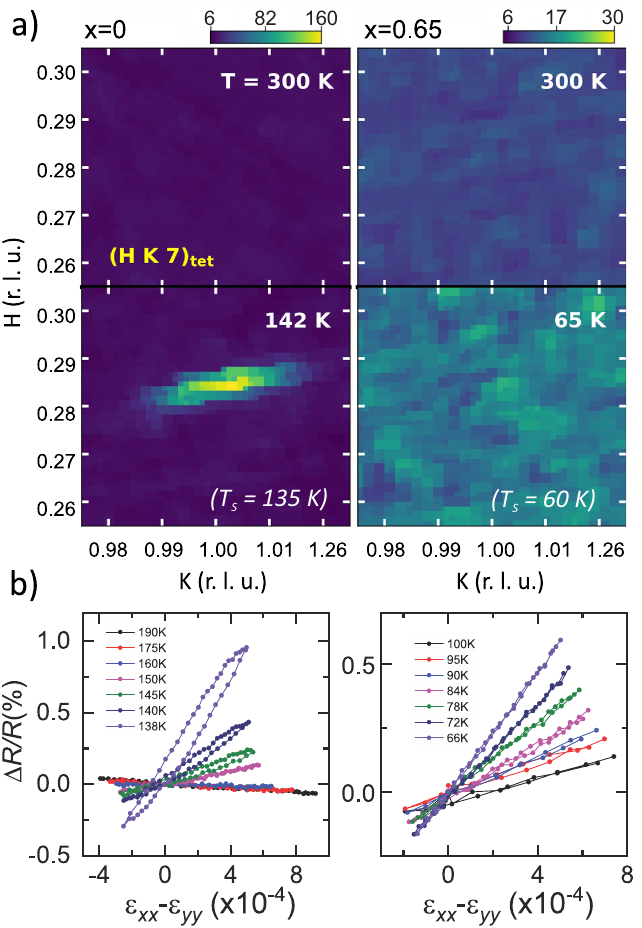}
    \caption{a.) ($H,K$) cuts through momentum space centered on $(0.28,1,7)_{tet}$ at room temperature and relative temperatures above the triclinic structural distortion for $x$=0 and $x$=0.65 Sr concentrations. Data taken from suplimentary figure S3 of \cite{Eckberg_nematic}. At room temperature for both concentrations no CDW is present. At $x$=0 the IC-CDW is observed at 142 K. At $x$=0.65, the IC-CDW is no longer observable above background levels. No other CDW peaks were detected in tetragonal phase for $x$=0.65 in all other regions of observed momentum space. X-ray data taken from supplementary figure S3 of \cite{Eckberg_nematic}. b.) Elastoresitivity ratios for $x$=0 and $x$=0.63 for temperatures just above the triclinic distortion. In $x$=0, the signal is hysteretic with strain, indicating nematic order. In $x$=0.63, the signal is reversible with strain. The observation of nematic order and the IC-CDW occur simultaneously for \BaSr.}
    \label{fig:No ICCDW}
\end{figure}

\begin{figure}
    \centering
    \includegraphics[width=0.47\textwidth]{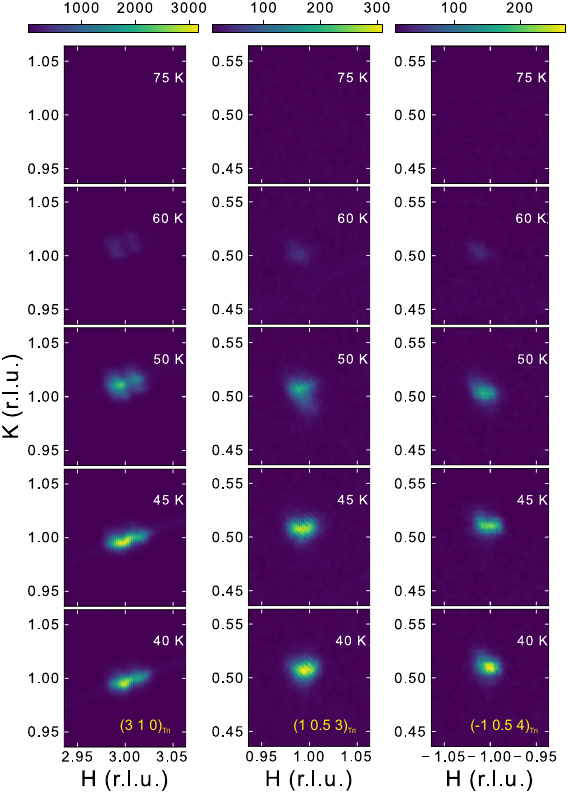}
    \caption{($H,K$) momentum space cuts in the triclinic space at $x$=0.65 for selected temperatures. The $Q_{tri}=0.5$ C-CDW2 observed via superstructure reflections at $(1,0.5,3)_{tri}$ and $(-1,0.5,4)_{tri}$. An example triclinic Bragg reflection onset is shown at $(3,1,0)_{tri}$. At 75 K, above the triclinic distortion, no scattering is found in these selected spaces. Intensity from each reflection is first observed starting at 60 K and continues to be observed down to 11 K (not shown).}
    \label{fig:CDW Late}
\end{figure}


Each charge order peak and structural transition is detectable by x-ray diffraction. 
At a Sr concentration of $x$=0.65, we report that there is no IC-CDW or any other charge order observed within the tetragonal phase, as presented in Fig.~\ref{fig:No ICCDW}. When the system undergoes the triclinic distortion, now at a reduced temperature of 60 K, only the C-CDW2 $Q=0.5$ order is observed (Fig.~\ref{fig:CDW Late}). Increasing Sr content slightly further to $x$=0.71 has previously shown a vanishing of the triclinic and C-CDW2 order along with an enhancement of $T_c$ up to 3.6 K \cite{Lee_Paper2,Eckberg_nematic}. The vanishing of the IC-CDW at $x$=0.65 in the tetragonal phase, along with the shift of commensurate order from C-CDW1 to C-CDW2 in the triclinic phase, indicates that the charge ordered phases of \BaSr~develop in a complex way and are not simply tuned to zero temperature.


The charge order of $x$=0.65 has quite different behavior than that of $x$=0. Pure \BaNi\ has a triclinic distortion at 135 K that coincides with the destruction of the $Q_{tet}=0.28$ incommensurate charge order and the emergence of a $Q_{tri}=0.33$ commensurate charge order \cite{Lee_BANI_CDW}. In the absence of a precursor incommensurate order for $x$=0.65, we observe at triclinic distortion at 60 K that, in this case, is accompanied by commensurate charge order with wavevector $(0,0.5,0)_{tri}$ (Fig.~\ref{fig:CDW Late}). 
The absence of incommensurate order in the tetragonal phase is likely associated with the absence of nematic order, as they are closely tied in \BaSr. Elastoresistance measurements for \BaNi~have shown a hysteresis as a function of applied strain along the B$_{1g}$ symmetry channel (Fig.~\ref{fig:No ICCDW}), which operates along equivalent crystallographic directions of the IC-CDW. At low Sr concentrations, the hysteresis in elastoresistance is thought to arise due to domain formation in an ordered nematic phase present in the system, and appears simultaneously with the onset of the bi-directional IC charge order \cite{Eckberg_nematic}. 

As Sr concentration increases, the window for nematic order above the triclinic distortion begins to narrow and vanishes around $x$=0.5 with a transition to electronically driven nematic fluctuations \cite{Eckberg_nematic}. Concurrently, the window of existence of the IC-CDW also narrows, and the IC-CDW has an intensity profile with temperature that scales with the hysteretic area profile seen in elastoresistance \cite{Eckberg_nematic}.
For $x$=0.65, we show that no IC-CDW or any other charge order exists above the triclinic distortion. A prior study has shown a completely reversible nematic susceptibility for this region of Sr concentration, and therefore, no signal of nematic order \cite{Eckberg_nematic}. We claim that charge order and nematic order not only scale together throughout \BaSr, but also mutually exist together throughout the entire phase diagram. Together with prior elastroresistivity measurements \cite{Eckberg_nematic}, our X-ray data (Fig.~\ref{fig:No ICCDW}) suggest there is an intimate tie between the two orders requiring further study to elucidate.


In conclusion, we report x-ray, transport, and magnetometry measurements showing an absence of long-range charge order in \BaSrLate~ within the tetragonal phase that is consistent with the absence of nematic order found previously. Together with the observation of a unique commensurate charge order with wavevector of $(0,0.5,0)_{tri}$ below the triclinic distortion, which evidences a complex evolution of charge-ordered phases with Sr substitution, 
these data help elucidate the rich evolution of electronic degrees of freedom in the \BaSr\ system.

X-ray experiments were supported by the U.S. Department of Energy, Office of Basic Energy Sciences grant no. DE-FG02-06ER46285 (PA), and the Institute for Complex Adaptive Matter (JC).
Low temperature measurements were supported by the National Science Foundation Grant no. DMR1905891 (JP), and JP and PA acknowledge the Gordon and Betty Moore Foundation's EPiQS Initiative through grant nos. GBMF9071 and GBMF9452, respectively. 

\bibliography{BSNA_CDW}

\end{document}